\documentclass[fleqn,usenatbib]{mnras}
\usepackage[T1]{fontenc}

\DeclareRobustCommand{\VAN}[3]{#2}
\let\VANthebibliography\thebibliography
\def\thebibliography{\DeclareRobustCommand{\VAN}[3]{##3}\VANthebibliography}

\usepackage{graphicx}	
\usepackage{amsmath}	
 \usepackage{amssymb}	
\usepackage{subfigure}
\usepackage{xcolor}
\usepackage{newtxtext,newtxmath}

 \newcommand{\kms}{km s$^{-1}$}

 \newcommand{\be}{\begin{equation}}
 \newcommand{\ee}{\end{equation}}
 \newcommand{\ba}{\begin{eqnarray}}
 \newcommand{\ea}{\end{eqnarray}}
 \newcommand{\bs}{\begin{subequations}}
 \newcommand{\es}{\end{subequations}}
 \newcommand{\s}{\nobreak\hspace{.11em}\nobreak}
 \definecolor{rkka}{RGB}{219,66,32}

\begin{document}
\title[Tidal decay of Kepler 1658 b]{On the orbital decay of the gas giant Kepler$\s$-$\s$1658$\s$b}

\author[Authors et al.]{
Adrian J. Barker$^{1}$\thanks{E-mail: A.J.Barker@leeds.ac.uk},
Michael Efroimsky$^{2}$\thanks{E-mail: Michael.Efroimsky@gmail.com},
Valeri V. Makarov$^{2}$\thanks{E-mail: Valeri.Makarov@gmail.com}
and
Dimitri Veras$^{3,4,5}$\thanks{E-mail: Dimitri.Veras@aya.yale.edu},
\\
$^{1}$Department of Applied Mathematics, School of Mathematics, University of Leeds, Leeds LS2 9JT\, UK\\
$^{2}$US Naval Observatory, Washington DC 23450 USA\\
$^{3}$Centre for Exoplanets and Habitability, University of Warwick, Coventry CV4 7AL\, UK \\
$^{4}$Centre for Space Domain Awareness, University of Warwick, Coventry CV4 7AL\, UK \\
$^{5}$Department of Physics, University of Warwick, Coventry CV4 7AL\, UK
}
\date{Accepted Nov 13 2023. Received Oct 27 2023; in original form Sept 22 2023}
\pubyear{2023}

\label{firstpage}
\pagerange{\pageref{firstpage}--\pageref{lastpage}}
\maketitle
\begin{abstract}
 The gas giant Kepler-1658b has been inferred to be spiralling into its sub-giant F-type host star Kepler-1658a (KOI-4). The measured rate of change of its orbital period is $\stackrel{\bf\centerdot}{\textstyle{P}}_{\rm orb}\,=\,-\,131^{+20}_{-22}\;\mbox{ms/yr}\s$, which can be explained by tidal dissipation in the star if its modified tidal quality factor is as low as $Q^{\,\prime}\approx 2.50\times {10}^{4}$. We explore whether this could plausibly be consistent with theoretical predictions based on applying up-to-date tidal theory in stellar models (varying stellar mass, age, and metallicity) consistent with our newly-derived observational constraints. In most of our models matching the combined constraints on the stellar effective temperature and radius, the dissipation in the star is far too weak, capable of providing $Q^{\,\prime}\gtrsim 10^9$, hence contributing negligibly to orbital evolution. Using only constraints on the stellar radius, efficient tidal dissipation sufficient to explain observations is possible due to inertial waves in the convective envelope during the sub-giant phase, providing $Q^{\,\prime}\sim 10^4$, but this period in the evolution is very short-lived (shorter than $10^2$ yrs in our models). We show that dissipation in the planet is capable of explaining the observed $\dot{P}_\mathrm{orb}$ only if the planet rotates non-synchronously. Tidally-induced pericentre precession is a viable explanation if the periastron argument is near $3\pi/2$ and the quadrupolar Love number is above 0.26. Further observations constraining the stellar and planetary properties in this system have the exciting potential to test tidal theories in stars and planets.
\end{abstract}


 \section{Introduction
   \label{Section_Motivation}
 }

The majority of the known exoplanet population will be destroyed through tidal engulfment into their parent stars. Although this process can occur along the main sequence only for planets within about 0.1~au \citep[e.g.][]{rasetal1996,BO2010,lai2012,weinberg,barker}, planets at separations as wide as 1-5~au will not survive the giant branch phases of stellar evolution \citep[e.g.][]{musvil2012,adablo2013}. Those planets which do survive the giant branch gauntlet are then assumed to play a vital role as dynamical drivers of white dwarf metal pollution \citep[e.g.][]{veras2016,veras2021}. These outcomes demonstrate the importance of understanding and constraining planet-star tidal interactions to a level which allows for accurate population synthesis investigations, as well as interpretations of noteworthy systems for which the stellar age has been precisely constrained. For example, the evolutionary history of the uniquely-located post-main sequence planets HD 203949~b \citep{cametal2019} and 8 Ursae Minoris~b \citep{honetal2023} cannot be determined without making assumptions about how efficiently tidal flows are dissipated.

One way to constrain efficiencies of tidal dissipation in stars (often quantified by modified tidal quality factors $Q^{\,\prime}$) is to measure the change in the orbital period of a tidally decaying exoplanet. However, even 30 years after the first confirmation of an exoplanet \citep{wolfra1992,wolszczan1994} such measurements are tentative, and have been undertaken for only a few systems \citep[e.g.][]{macetal2018,yeeetal2020,debetal2023,haretal2023}.

One of these systems is Kepler-1658, which is noteworthy partially for having properties indicating that it had recently evolved off the main-sequence. The gas giant Kepler-1658 b is a massive planet with a comparable size to Jupiter (mass $M_{\rm p}\s=\s 5.88M_J$ and radius $R_{\rm p}\s=\s 1.07R_J$, with Jupiter's mass $M_J$ and radius $R_J$) observed to orbit an evolved F-type star (with stellar mass $M\approx 1.6M_\odot$) likely to be in the sub-giant phase shortly after the main sequence \citep{Chontos2019}. The planet has a short orbital period $P_{\rm orb} = 3.85$ d, which is very close to synchronism with the stellar rotation period $P_{\rm rot} \approx 4$ d, and the orbit has been inferred to be shrinking at the rate $\stackrel{\bf\centerdot}{\textstyle{P}}_{\rm orb}\,=\,-\,131^{+20}_{-22}\;\mbox{ms/yr}$ =$\,-\,( 415 ^{+63}_{-70})\times10^{-11}\;\mbox{s/s}$, corresponding to a characteristic inspiral time for orbital decay of approximately 2.5 Myr \citep{Vissapragada}. This is much shorter than the estimated age of the star, so we are fortunate to observe the system in its current state. \citet{Vissapragada} deduced this rate of inspiral to imply efficient tidal dissipation inside the star, corresponding to a modified quality factor $\s Q^{\,\prime}=2.50^{+0.85}_{-0.62}\times 10^{4}\s$. They proposed that this value agrees with theory, and hypothesised that the decay could be explained by efficient tidal excitation and dissipation of inertial waves in the convective envelope of the star when it passes through the sub-giant phase \citep[based on Figure 6 of][hereafter B20]{barker}.

 In this paper we revisit the fascinating Kepler 1658 system to determine whether its inferred $\dot{P}_{\rm orb}$ could be explained by stellar tidal theory. To do so, we first derive new constraints on the properties of the star (mass, radius, age, metallicity and rotation rate) using the Gaia catalogue coupled with stellar evolutionary models. We then compute stellar tidal dissipation rates theoretically in models matching these observational constraints (following B20). We find $\s Q^{\,\prime}\sim 10^4$ is possible due to dissipation of inertial waves in the convective envelope, but there is a large uncertainty due to the rapid variation in stellar properties as the star evolves off the main sequence through this phase. We find that most models matching the combined constraints on stellar effective temperature and radius predict $\s Q^{\,\prime}\gtrsim 10^9$ instead, which would result in negligible orbital evolution of the planet. We discuss the implications of our results, and we also show that the orbital evolution cannot be explained by planetary tides unless there is sustained non-synchronous rotation. An alternative explanation for the observed $\stackrel{\bf\centerdot}{\textstyle{P}}_{\rm orb}$ is presented, which is related to the tidal deformation of a synchronised planet and the subsequent precession of the line of apsides. This scenario requires a finite orbital eccentricity, an orbital configuration with the line of apsides roughly aligned with the line of sight, and a quadrupolar Love number above a certain -- but reasonable -- value.

 \section{Stellar parameters and rate of rotation}
  \label{star.sec}

 Kepler-1658 (KOI-4) is an evolved star that is more massive than the Sun, and whose evolutionary status and age can be estimated using available stellar evolution models. We employ the PARSEC \citep
  {2012MNRAS.427..127B, 2014MNRAS.445.4287T} evolution models, which are best suited for the accurately-determined photometric magnitudes and parallaxes in the Gaia DR3 catalog \citep{2021A&A...649A...1G,2016A&A...595A...1G} and have been validated on open clusters including the Hyades \citep[e.g.,][]{2021A&A...649A...6G}.  We find the following data in the Gaia catalogue: broadband magnitude $G=11.417\pm 0.003$, blue magnitude $G_{\rm BP}=11.694\pm0.003$, red magnitude $G_{\rm RP}=10.981\pm0.004$, parallax $\varpi=1.226\pm 0.017$ mas. Moderate values of the metadata parameters  \texttt{phot\_bp\_rp\_excess\_factor}$=1.19$ and \texttt{ruwe}$=1.15$, as well as a low value of \texttt{ipd\_gof\_harmonic\_amplitude} indicate a stable, unperturbed astrometric and photometric solution \citep{2021A&A...649A...5F}.
  The nearest neighbour $3.37\arcsec$ away is an unrelated star that is fainter by 4.6 mag---unlikely to perturb the data for Kepler-1658. 
  
\begin{figure}
    \includegraphics[width=0.47 \textwidth]{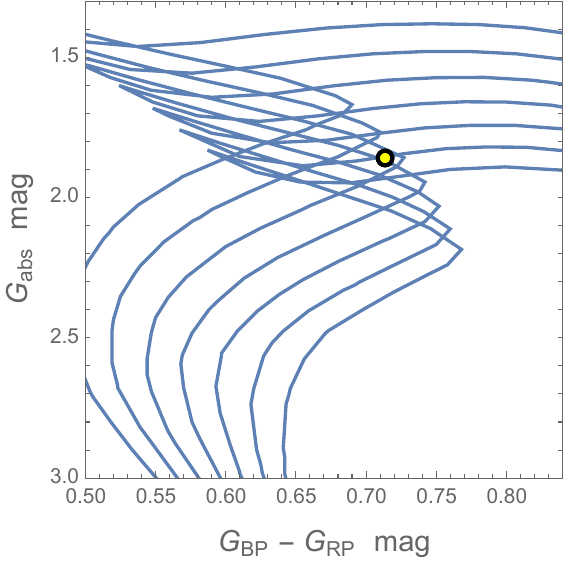}
    \caption{Colour-magnitude diagram for Kepler-1658 and selected isochrones from PARSEC stellar evolution models utilising Gaia DR3 data. The location of the star is marked with a yellow dot.
    The isochrones are computed for a grid of ages 1.5, 1.6, 1.7, 1.8, 1.9, 2.0, and 2.1 Gyr, from left to right at the bottom. Other relevant parameters for the stellar models are described in the text.}
    \label{cmd.fig}
\end{figure}

  Figure \ref{cmd.fig} shows the corresponding location of the target
  on the colour-magnitude diagram. Note that the observational errors are insignificant, which explains the absence of error bars. A grid of seven isochrones from 1.5 through 2.1 Gyr is shown with solid lines.
  The main uncertainty in matching the observational data with stellar isochrones comes from the imprecise input model parameters. For this computation, we assume
  an interstellar extinction of $A_V=0.378$ mag from \citet{2018ApJ...866...99B}, and a metallicity value [Fe/H]$=-0.247$ from \citet{2018AJ....155...68W}. The substellar metallicity places the star in an area where the model isochrones are tangled, which precludes an unambiguous determination of model parameters. We note some dispersion of input parameter values in the literature. For example, a smaller value for
  $A_V$ of 0.22 mag is estimated by \citet{2017AJ....154..259S}. Alternative determinations of metallicity include [Fe/H]$=-0.16\pm 0.15$ based on pre-Gaia data \citep{2014ApJS..211....2H}, [Fe/H]$=-0.27$ from
  the California-Kepler survey \citep{2022AJ....163..179P}, and [M/H]$=-0.099\pm 0.024$ from APOGEE-2 DR16 \citep{2020yCat.3284....0M}. A higher metallicity shifts the isochrones mostly to redder colours, but it
  hardly helps to resolve the ambiguity of age. With the assumed parameters in Figure \ref{cmd.fig}, the closest isochrone provides age $=1.8$ Gyr, mass $M=1.62$ $M_{\odot}$, $T_{\rm eff}=6628$ K, $\log \,g=3.64$.
  However, a different isochrone from the models comes to the observed position within 0.02 mag: age $=1.7$ Gyr, mass $M=1.65$ $M_{\odot}$, $T_{\rm eff}=6577$ K, $\log \,g=3.63$. The available data do not
  allow us to discriminate between these models. Our estimates for $T_{\rm eff}$ are significantly higher than some of the values previously given in the literature \citep[e.g.,][]{2012yCatp038048601B, 2018ApJ...866...99B, 2020yCat.3284....0M}, but are consistent with or lower than others \citep[e.g.,][]{2021ApJS..255...17S}. Our $\log \,g$ estimates are quite close to the values quoted in the literature. The previously estimated masses, in sync with the effective temperature, tend to be lower than our values, with the exception of \citet{2021ApJS..255...17S} where $M=1.781^{+0.064}_{-0.139}$ $M_{\odot}$ is provided.

Ground-based spectroscopic determinations of $v\,\sin i$ indicate a high rate of rotation. Some of the quoted numbers are
 41.2 \kms\ \citep{2012yCatp038048601B}, 36.9 \kms\ \citep{2020ApJ...889...54M}, and 38.0 \kms\ \citep{2022AJ....163..179P}.
 We should also consider the Gaia DR3 determination of $V_{\rm broad}=34.5\pm 8.4$ \kms. This parameter is statistically close to $v\,\sin i$ in the appropriate range of $T_{\rm eff}$ \citep{2022arXiv220610986F}. If we assume this latter value for Kepler-1658, then, by scaling with the known surface velocity of solar rotation and using the Carrington period, we obtain a rotation period of $P_{\rm rot}\simeq 4.4\pm 0.4$ d for an assumed radius of $3\;R_{\rm sun}$. A higher value of $v\,\sin i=38.0$ \kms\ corresponds to a period of 4.0 d. We note that that these estimates are close to the orbital period of planet b (3.85 d), and, given the large uncertainty in Kepler-1658's radius, it is possible that the star has (or at least the surface layers have) been spun up by the planet and it is now rotating synchronously, or close to this state.

\section{Tidal parameters: Notation\label{Section_notation}}

 The mass and radius of the star are denoted with $M$ and $R$; those of the planet with $M_{\rm p}$ and $R_{\rm p}$. For a tidally perturbed body, the Love numbers $k_l$, phase lags $\epsilon_l$, and quality factors $Q_l\equiv|\sin\epsilon_l|^{-1}$ are functions of the Fourier components of the tide, i.e., of the tidal frequencies $\omega_{lmpq}$. An $l$-degree {\it{quality function}}, sometimes referred to by the Danish word {\it{kvalitet}}, is defined as
    \citep[cf. ][]{2018ApJ...857..142M}
   \ba
  \nonumber
 K_l(\omega_{lmpq})&\equiv&
k_l(\omega_{lmpq})\,\sin\epsilon_l(\omega_{lmpq})  \label{1}\\ &=&\frac{k_l(\omega_{lmpq})} {Q_l(\omega_{lmpq})}\;\mbox{Sign}\,\omega_{lmpq}\;\,,\quad
 \nonumber
 \ea
 The quadrupolar quality factor is conventionally written with no subscript:  $Q\equiv Q_2$.
 It is also common to employ the {\it{modified quality factor}} introduced as
  \ba
  Q^{\,\prime}\,\equiv\,\frac{3}{2}\;\frac{1}{|K_2|}\,=\,\frac{3}{2}\;\frac{Q}{|k_2|}\;\,.
  \label{3}
  \ea

 \section{Tidal dissipation in the star\label{Section_dissipation_in_star}}
 \vspace{1mm}

\begin{figure*}
\centering
\vspace{-0.35cm}
 \includegraphics[trim=0cm 0cm 32.5cm 0cm,clip=true,width=1.13\textwidth]{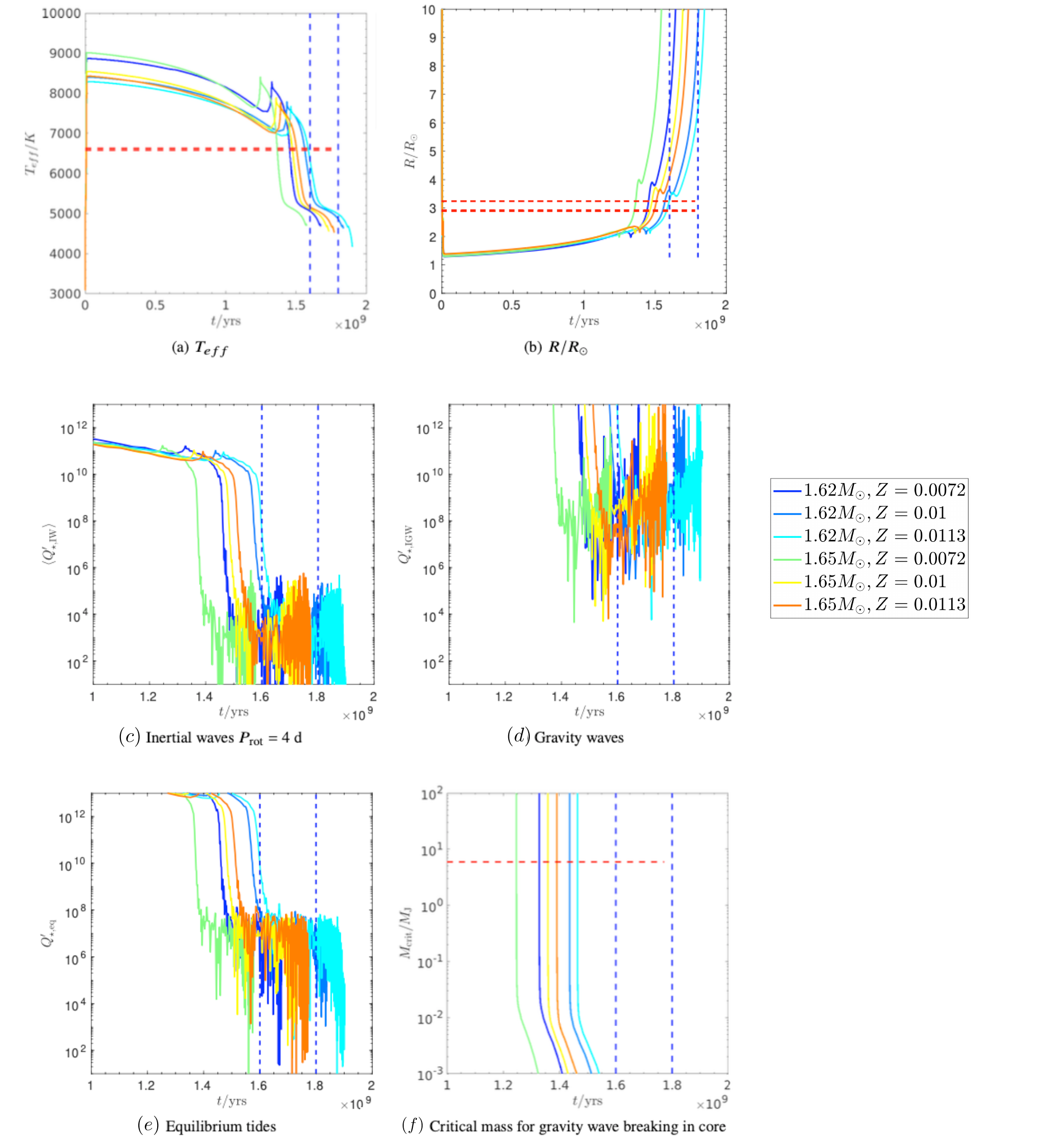}
  \caption{Stellar properties and tidal quality factors computed using default MESA parameters for the initial masses and metallicities specified in the legend, assuming $P_\mathrm{rot}=4$ d.  Panels (a) and (b) show the stellar effective temperature and normalised radius, over-plotting the constraints from section 2 as horizontal red dashed lines and a range of ages approximately consistent with PARSEC models by vertical dashed lines. Panels (c), (d) and (e) show modified tidal quality factors $Q^{\,\prime}$ for inertial waves, gravity waves and equilibrium tides, respectively. Panel (f) indicates the critical planetary mass required for wave breaking to be predicted in the stellar core. When the core becomes radiative, the critical mass for wave breaking is easily exceeded by the planet (i.e.~horizontal red dashed line, indicating Kepler 1658 b is above the critical mass lines). In all models as the star evolves off the main sequence, there is a rapid change in stellar properties and tidal dissipation rates. During this phase inertial waves are predicted to be the dominant tidal mechanism for such a rapidly rotating star (note that $\langle Q'_\mathrm{IW}\rangle\propto P_\mathrm{rot}^2$).}
  \label{AB1}
\end{figure*}

\subsection{Modified quality factor $Q^\prime$ of the star}
 \label{Qstar}
We now turn to build interior models of Kepler-1658 that match the observational constraints from Section 2, and perform calculations to determine tidal dissipation rates predicted by several mechanisms based on the latest theoretical expectations. To do so, we first construct stellar models based on the {\sc MESA} code with default parameters for all physical quantities except initial mass and metallicity \citep{Paxton2011,Paxton2013,Paxton2015,Paxton2018,Paxton2019,Jermyn2022}. We also computed models using the parameters from MIST \citep{MIST02016,MIST12016}, but did not find substantial quantitative differences. These computations provide us with 1D evolutionary models of Kepler-1658, which generate, for each time (age), profiles for the density $\rho(r)$, pressure $p(r)$, gravitational acceleration $g(r)$, Brunt-V\"{a}is\"{a}ll\"{a} (buoyancy) frequency $N(r)$, as well as the convective velocity $v_c(r)$ and mixing-length $\ell_c(r)$ in convective regions. Using these profiles, we can compute the linear tidal response in the star following B20. In particular, we compute the equilibrium (non-wavelike) tide in the convective envelope by solving equations (12) and (17) from Section 2 in B20 for the tidal component with harmonic degree and azimuthal wavenumber $l=m=2$, which determines the resulting (irrotational) displacement $\boldsymbol{\xi}_{\mathrm{nw}}(\boldsymbol{r},t)$ in this model. The equilibrium tide is damped by its interaction with turbulent convection, a process we model by assuming it can be described with an isotropic kinematic viscosity $\nu_E$ that is a function of radius, and which behaves in otherwise the same way as the (negligibly small) microscopic kinematic viscosity of the fluid in damping the tide. We use the numerical coefficients in equation (27) of B20 for $\nu_E$, obtained using the extensive suite of hydrodynamic simulations of \citet{DBJ2020}. We then perform the integral over radius in the stellar model using equation (20) from B20 to find the total ``viscous'' dissipation rate $D_\nu$ for a given tidal component and stellar model. This is converted to a stellar tidal quality factor $Q^{\,\prime}$  using:
\begin{equation}
    Q^{\,\prime}=\frac{3(2l+1)R^{2l+1}}{16\pi G}\frac{|\omega||A|^2}{D_\nu}\;\,,
\end{equation}
where $\omega=2\pi /P_{\mathrm{tide}}=\omega_{lmpq}\,$, $\s G$ is the gravitational constant, and $a$ is the semi-major axis. The quantity $A \propto (M_{\rm p}/M)(R/a)^3$ is the amplitude of the tidal perturbation, so that the ratio $ D_{\nu}/|A|^2$ and hence $Q^{\,\prime}$ (hereafter with subscript eq to denote the contribution from equilibrium tides) is independent of the tidal amplitude.

 We also compute tidal dissipation rates from wavelike (dynamical) tides in both convective and radiative regions. In radiative regions,
 the response consists of tidally-excited internal gravity (or gravito-inertial) waves, and we compute the resulting dissipation by assuming
 these waves to be launched adiabatically as travelling waves from the radiative/convective interface (with the envelope; excitation from the
 convective core is much weaker in this star) and fully damped before they can reflect from the inner convective core to set up a standing mode. To do so, we employ equation (41) of B20 (and surrounding formalism) based on applying the ideas of \citet{Z1975} and \citet{GD981998}, to compute $Q^{\,\prime}$ corresponding to gravity waves (hereafter with subscript IGW). This is justified if the waves have large enough amplitudes to break or if they are damped by radiative diffusion or absorbed in a critical layer \citep[if the star rotates differentially, at this location the angular phase velocity of the wave matches the local stellar rotation and we expect significant wave absorption;][]{BO2010,B2011,Su2020,Guo2023}. Wave breaking is less likely in this star on the main-sequence than in a solar-type star with a radiative core, where the waves can attain large amplitudes due to geometrical focusing \citep{GD981998,OL2007,BO2010,B2011,Guo2023}. It is likely when the star evolves through the sub-giant phase and develops a radiative core at the centre however. In any case, this estimate of $Q^{\s\prime}_\mathrm{IGW}$ provides a typical magnitude of gravity wave damping. If waves are more weakly damped, larger tidal dissipation rates are possible if the tidal frequency becomes resonant with a global g-mode \citep[due to the larger amplitude tidal response attained; this could in principle be maintained if we invoke a resonance locking scenario;][]{Ma2021}. Note that in the fully-damped approximation,
 $ Q^{\,\prime}_{ \mathrm{IGW} } \propto P_{\mathrm{tide}}^{8/3}$ (a strong dependence!).

In convection zones of rotating stars in which convective motions are efficient at homogenising the entropy, and can enforce an approximately adiabatic stratification profile, the only wavelike motions that can occur (for low frequencies, in the absence of magnetic fields) are inertial waves. These waves are of low frequency (relative to the stellar dynamical frequency $\sqrt{GM/R^3}$) and are restored by Coriolis forces, being (linearly) excited only when tidal frequencies satisfy $|\omega|\leq 2|\Omega|$, where $\Omega=2\pi/P_\mathrm{rot}$ is the stellar rotational angular frequency. When these waves are excited they can provide a substantial enhancement of tidal dissipation rates in a strongly frequency-dependent manner \citep[e.g.][]{OL2007}. We calculate the wavelike tide by applying the frequency-averaged formalism of \citet{O2013}, which computes the energy transfer into inertial waves following an initial impulsive tidal forcing. This represents a ``typical level of dissipation" of inertial waves when they are excited, and is convenient to calculate because it only requires the solution of an ODE in radius (rather than a coupled system of 2D PDEs in the meridional plane to solve for the wavelike response directly). In our models we compute $Q^{\,\prime}$ for inertial waves (hereafter with subscript IW and with angled brackets $\langle \cdot\rangle$ to indicate it is based on the frequency-averaged dissipation) using equation (30) of B20. This calculation provides a typical magnitude for the resulting dissipation that ignores much of the complicated (and uncertain) frequency-dependence from linear theory. Our approach builds on the application of two-layer piece-wise homogeneous stellar models used by \citet{M2015}, \citet{Gallet2017}, and many others, to fully account for the realistic structure of the star. The resulting $\langle Q^{\,\prime}_\mathrm{IW}\rangle \propto P_\mathrm{rot}^2$, and thus is more efficient for more rapid rotation.

In our stellar models, at every time (age) $t$ for which we output the radial profiles, we compute each of $Q^{\,\prime}_{\mathrm{eq}}\,$, $\s Q^{\,\prime}_{\mathrm{IGW}}$ and $\langle Q^{\,\prime}_{\mathrm{IW}}\rangle$ as described. We set $P_\mathrm{rot}=4$ d, $P_\mathrm{orb}=2\pi/n=3.85$ d, such that $P_{\mathrm{tide}}=2\pi/\omega$, where $\omega=2(n-\Omega)$, $M_{\rm p}=5.88 M_J$ and use a variety of models varying the stellar mass $M\in\{ 1.62,\,1.65\} M_\odot$ and metallicity $Z\in\{ 0.0072,\,0.01,\,0.0113\}$. Note that the stellar rotation period is uncertain.

Results are shown in Figure~\ref{AB1}, where we have used default MESA parameters except those specified in the legend for initial mass and metallicity (MIST input files produce similar results). Panels (a) and (b) show the effective temperature $T_{\rm eff}$ and solar-normalised stellar radius as a function of age in years, respectively, highlighting observational constraints as red horizontal dashed lines, and a range of ages approximately consistent with PARSEC models by vertical dashed lines. The only evolutionary stages consistent with observed values are when these stars evolve off the main-sequence, when they undergo a rapid drop in $T_{\rm eff}$ and increase in $R$. This occurs for a particular age that depends on initial mass and metallicity, occurring for earlier ages for more massive stars ($M=1.65M_\odot$) and those with lower metallicities ($Z=0.0072$). Given the uncertainties in stellar age, we notice that each model passes through a value of $T_{\rm eff}$ and $R$ consistent with observations for some age in the range $1.4-1.6$ Gyr (for lower mass models this occurs for even later ages). At this phase in the evolution the star is predicted to have a radiative core. As such, the planet is predicted to excite gravity waves that attain large amplitudes near the centre sufficient to cause wave breaking. The critical planetary mass required for wave breaking is plotted in panel (f), and we notice that this is typically much lower than the red dashed line (corresponding with $M_p$) for the ages inferred from panels (a) and (b), such that wave breaking in the stellar core is predicted. This is likely to justify our assumption that gravity waves are fully-damped in the core. The resulting $Q^{\,\prime}_{\mathrm{IGW}}$ is given in panel (e). For our adopted $P_{\mathrm{rot}}$ (and hence $P_{\mathrm{tide}}$), we find $Q^{\,\prime}_{\mathrm{IGW}}\gtrsim 10^8$. This suggests gravity wave dissipation is unlikely to explain the inferred $\dot{P}_\mathrm{orb}$.

In panel (f), we show $Q^{\,\prime}_{\mathrm{eq}}$ from equilibrium tide damping in the convective envelope. For ages consistent with observations in panels (a) and (b), we find $Q^{\,\prime}_{\mathrm{eq}}\gtrsim 10^7$. Note that our choice of $P_{\mathrm{rot}}$ means that $P_{\mathrm{tide}}$ is not so short that the frequency-reduction of $\nu_E$ for fast tides substantially inhibits this mechanism. However, this is still insufficient to explain the observationally-inferred value.

Finally, we show inertial wave dissipation $\langle Q^{\,\prime}_{\mathrm{IW}}\rangle$ in panel (c). This mechanism is predicted to be the most efficient one during this phase. We find for ages consistent with $T_{\rm eff}$ matching observations in panel (a), that $\langle Q^{\,\prime}_{\mathrm{IW}}\rangle$ rapidly falls such that it can attain values as low as $10^2-10^4$. Thus, our models have shown that values for $Q^{\,\prime}$ consistent with observations are theoretically possible for inertial waves. A caveat is that while the star is undergoing the rapid evolution in $T_{\rm eff}$ through the observed value, $Q^{\,\prime}$ due to inertial waves rapidly falls from values larger than $10^{10}$ such that it is difficult to make a robust prediction of $Q^{\,\prime}$ unless we are convinced by our current understanding of stellar models and the observational constraints. It is possible (indeed likely) that the $Q^{\,\prime}$ resulting from inertial waves at the specific tidal period could differ from the frequency-averaged prediction computed here by an uncertain amount (potentially by several orders of magnitude). Nevertheless, our results are consistent with inertial waves in the convective envelope being the most effective tidal dissipation mechanism, and values consistent with observations are certainly attainable (in all of the models with different $M$ and $Z$ that we have studied) depending strongly on the phase of stellar evolution in which the planet is observed.

\begin{table}
\begin{center}
\begin{tabular}{ c|c|c|c }
$M/M_\odot$ & $Z$ & $T_{\rm eff}\in[6576,6629]K$ & $R\in [2.6,3.9] R_\odot$ \\
\hline
1.62 & 0.0072 & -                 &  $[0.79,  \, 2.08\times 10^{10}]$ \\
1.62 & 0.01   & $3.25\times 10^9$ &  $[0.57,  \, 2.64\times 10^{10}]$ \\
1.62 & 0.0113 & $3.52\times 10^9$ &  $[9.83\times 10^3, \, 2.04\times 10^{10}]$ \\
1.65 & 0.0072 & $3.60\times 10^9$ &  $[405.8, \, 2.21\times 10^{10}]$ \\
1.65 & 0.01   & $3.38\times 10^9$ &  $[52.6,  \, 2.19\times 10^{10}]$ \\
1.65 & 0.0113 & $2.83\times 10^9$ &  $[0.79,  \, 2.08\times 10^{10}]$ \\
\end{tabular}
\caption{Table indicating predicted ranges of $\langle Q^{\,\prime}_\mathrm{IW}\rangle$ for which $T_{\rm eff}$ or $R$ from our models pass through observational constraints. The dash in the first row indicates that no snapshot computed passes through this phase due to limited time resolution of MESA output files.}
\label{tableQconstraints}
\end{center}
\end{table}

The most likely value of $Q^{\,\prime}$ resulting from inertial wave dissipation is uncertain, even assuming $\langle Q_{\mathrm{IW}}^{\,\prime}\rangle$ perfectly represents their dissipation. This is because of rapid changes in stellar properties and values of $\langle Q_{\mathrm{IW}}^{\,\prime}\rangle$ during this phase. Values as small as $10^4$ --- approximately consistent with observations --- or as large as $10^{10}$ are possible depending on whether the star is observed towards the end or start of this rapid evolutionary phase from the main-sequence through the sub-giant phase. The period in the evolution in which $\langle Q_{\mathrm{IW}}^{\,\prime}\rangle\sim 10^4$ and so can explain observations is very short-lived in our models, being shorter than $10^2$ yrs. If we require the effective temperatures between $6576$ and $6629$ K, we would predict values in excess of $Q^{\,\prime}\approx 10^9$ for $P_{\mathrm{rot}}=4$ d. If we require the radius to be between $2.6 R_\odot$ and $3.9R_\odot$ then we can obtain values of $Q^{\,\prime}$ as small as $O(1)$ to $10^3$ or as large as $10^{10}$. These values are summarised in Table~\ref{tableQconstraints}.  We therefore conclude that values consistent with observations, where $Q^{\,\prime}\approx 10^4$, are certainly attainable due to inertial waves. However, combining our $T_{\rm eff}$ and $R$ constraints (where the former gives the tightest range) in Table~\ref{tableQconstraints}, we find values closer to $10^9\;$:
\ba
Q^{\,\prime}\simeq 3\times 10^9\;\;,\qquad |K_2|\,=\,\frac{3}{2}\,\frac{1}{Q^{\,\prime}}\,\simeq\,5\times 10^{-10}.
\label{stellarQ}
\ea
We show in the next section that these values -- if they are the appropriate ones -- would predict negligible tidal evolution of the orbit. 

We also point out an alternative possibility here. If the star is in fact synchronised with the planet's orbit, and if tidal dissipation is efficient enough to continue to rapidly synchronise the spin and orbit, we would predict the orbit to decay on the magnetic braking timescale \citep[e.g.][]{BO2009,Damiani2015}. For an F-star like Kepler 1658, we estimate this timescale to be longer than $0.2$ Gyr, which is far too long to explain the observed $\dot{P}_\mathrm{orb}$.

\subsection{Orbital evolution due to tides in the star}

 From $P_{\rm orb}=2\pi/n$ and
 $n=\sqrt{G(M+M_{\rm p})\,a^{-3}\,}$,
 we obtain:
 \ba
 \stackrel{\bf\centerdot}{P\,}_{\rm orb}\;
 =\,\frac{3\,\pi}{n\,a}\,\dot{a}\;\;.
 \label{period_rate}
 \ea
 With a small orbital eccentricity $e$ and a nearly aligned orbit (with stellar spin-orbit angle $i$), an $e^2$-approximation for the primary's and secondary's contributions to the tidal migration rate $da/dt$ can be found in e.g.~\citet[Section 4.2]{Boue}. The input from a non-synchronous primary (the star) is
 \begin{align}
 \nonumber
 \left(\frac{da}{dt}\right)^{\rm (star)}=
 -\,3\;n\,a\;\frac{M_{\rm p}}{M\;}\,\left(\frac{R}{a}\right)^{\textstyle{^5}}\,K_{2}(2n-2\dot{\theta})\;\qquad
 ~\\
 \nonumber
 -\,\frac{3}{8}\, n\s a\s e^2\s\frac{\;\,M_{\rm p}}{M}\s\left(\frac{R}{a}\right)^{\textstyle{^{5}}}\s\left[
 -\s 40\, K_{2}(2n-2\dot{\theta})\s +\s 6\s K_2(n)\s\right.
 ~\\
 \left.  \label{comparison}
 +\s K_2(n-2\dot{\theta})\s +\s 147\s K_2(3n-2\dot{\theta})
 \s\right]
 ~\\
 \nonumber
 -\,\frac{3}{4}\;a\;n\,\left(\frac{R}{a}\right)^7\,\frac{M_{\rm p}}{M}\;
 \left[\,5\;K_3(3n-3\dot{\theta})\;+\;K_3(n-\dot{\theta})\,\right]
 \\
 \nonumber
  +\;O(i^2)\;+\;O(e^4) ~~_{\textstyle{_{\textstyle ,}}}
 \end{align}
 $\theta$ and $\dot{\theta}$ being the rotation angle and rotation rate of the star, and $i$ being the stellar obliquity on the orbital plane. In this expression, we have retained degree-3 terms of order $e^0$, because in tight systems they may be comparable to the quadrupolar $e^2$ terms. To leading order these are independent of $i$.

 For an estimate, we combine equation (\ref{period_rate}) with the first line of (\ref{comparison}):
 \ba
 \stackrel{\bf\centerdot}{P\,}_{\rm orb}^{\rm{(star)}}\;
 \approx\,-\,9\s\pi\,
 \frac{M_{\rm p}}{M\;}\,\left(\frac{R}{a}\right)^{\textstyle{^5}}\,K_{2}(2n-2\dot{\theta})
 \;\;.
 \label{equation}
 \ea
 The insertion of the ``most likely" value (\ref{stellarQ}) obtained for $K_2$ makes ${\dot{P}\,}^{\rm (star)}_{\rm orb}\approx 2.43\times 10^{-16}$ s/s = $76.7\times 10^{-7}$ ms/yr, which is more than seven orders of magnitude lower than the observed rate. This indicates that tidal dissipation in the non-synchronous star cannot produce the observed orbital shrinking, unless there exists additional physical factors boosting the dissipation in the star by more than seven orders of magnitude. This could happen if the stellar temperature or radius happen to be sufficiently poorly constrained that the star is really better represented by models that predict by $Q^{\,\prime} \sim 10^4$.

If the star happens to be synchronised with the planet, its effect on the orbital evolution decreases further by several orders of magnitude, because in this case only the $e^2$-order contribution (and $e\approx 0.0628$) will survive in expression (\ref{comparison}).

 \section{Tidal dissipation in a synchronised planet}

 The contribution from a synchronised secondary to $da/dt$ looks like the expression in (\ref{comparison}), though we now have to interchange $M$ with $M_{\rm p}$, and substitute $R$ with $R_{\rm p}$. We also have to substitute the stellar tidal frequencies $\omega_{lmpq}$ and quality function $K_2(\omega_{lmpq})$ with their planetary counterparts $\omega^{\rm(planet)}_{lmpq}$ and $K_{2,p}(\omega^{\rm (planet)}_{lmpq})\s$.

 With $\theta_{\rm p}$ and $\dot\theta_{\rm p}$ the planet's rotation angle and rate, synchronism implies $\dot{\theta}_{\rm p} = n\,$; \,and the resulting expression reduces to
 \begin{align}
 \nonumber
 \left(\frac{da}{dt}\right)^{\rm (planet)}_{\rm (synchr)} =\;
 -\;57\;a\,n\,e^2\,\left(\frac{R_{\rm p}}{a}\right)^{\textstyle{^{5}}}\frac{\;M\;}{M_{\rm p}}\;K_{2,p}(n)
  ~\\
 +\;O(i_p^{\,2})\;+\;O(e^4) \label{49}
 \,\;,
 \end{align}
 $i_{\rm p}$ being the planet's obliquity.   This expression is equivalent to
 \ba
 \nonumber
 \stackrel{\bf\centerdot}{P\,}_{\rm orb}^{\rm (planet)}
 \hspace{-0.3cm}&=&\hspace{-0.3cm}\frac{3\s\pi}{n\s a}\,\dot{a}^{\rm (planet)}_{\rm (synchr)}
 ~\\
 \hspace{-0.3cm}&=&\hspace{-0.3cm}
 -\;171\s\pi\,e^2\,\left(\frac{R_{\rm p}}{a}\right)^{\textstyle{^{5}}}\frac{\;M\;}{M_{\rm p}}\;K_{2,p}(n)\,+\,O(i_p^{\,2})\,+\,O(e^4)\;,\quad
 \label{pdot}
 \ea
 and can also be cast as
 \ba
 K_{2,p}(n)\s\approx\;-\;1.86\times 10^{-3}\,\frac{\;M_{\rm p}}{\;M\;}\;\left(\frac{a}{R_{\rm p}}\right)^{\textstyle{^{5}}}\;\frac{\stackrel{\bf\centerdot}{P\,}_{\rm orb}^{\rm (planet)}}{e^2}\;\,.
 \label{}
 \ea
 Holding the hypothetically synchronised planet solely responsible for the registered tidal decay, we identify rate (\ref{pdot}) with the actual measured rate: $\s\stackrel{\bf\centerdot}{P\,}_{\rm orb}^{\rm (planet)} = \s{\dot P}_\mathrm{orb} =\s-\, 4.15\times 10^{-9}$ s/s\,. The insertion of this value, along with the known values $\,e=6.28\times 10^{-2}\s$, $\,R_{\rm p}=7.48\times 10^7$~m\s, $\,a=8.14\times 10^{9}$~m\s, and ${M_{\rm p}}/M\s=\s 3.87\times 10^{-3}\s$, into the above expression entails, for the planet:
 \ba
 K_{2,p}(n)\s\approx\,1.16\times 10^{-1}\quad\mbox{and}\quad Q^{\,\prime}_{\rm p}=\frac{3}{2}\,\frac{1}{|K_{2,p}|}\,\approx\,1.29\times 10^1\,\;.
 \label{}
 \ea
These values do not look realistic in the light of our present understanding of planets' structure.
 
 While the above consideration is valid for a single-point estimation of the observed decay rate $da/dt$, the long-term evolution of the planet's orbit involves also the
 rate of eccentricity decay, because equation (\ref{49}) includes $e^2$. The pair of differential equations for $da/dt$ and $de/dt$ should be solved simultaneously as a system, to obtain the correct result \citep[see][using the constant time-lag model]{BO2009}. For most of the close exoplanets known today, the rate of eccentricity decay is much higher than the rate of semimajor axis decay, to the effect that integrating the equation
  \ba
 ~\left(\frac{de}{dt}\right)^{\rm (planet)}_{\rm (synchr)}&=&
  -\,\frac{21}{2}\,n\,e\,\frac{M\,}{\,M_{\rm p}}\,\left(\frac{R_{\rm p}}{a}\right)^{\textstyle{^{5}}}\,
 K_{2,p}(n)
 \label{de} \\
 \nonumber
 &\,&\hspace{2cm}+\,O({i^{\,\prime\;}}^2)  \,+\,O(e^2)
 \ea
 separately and assuming a constant $a$ yields a reasonably accurate result \textit{for small $e$}. For the same reason, a separate integration of the $da/dt$ equation produces a completely misleading result. The interesting consequence for this study is that the estimated $K_{2,p}(n)$ can be used to compute the characteristic $e$-folding time $e/(de/dt)$, giving $81$ Kyr with the same parameters as used above. Thus, tidal circularisation is a very rapid process, and any remnant eccentricity should have already been damped by planetary tides.
 
 With aid of equation (\ref{pdot}),  equation (\ref{de}) for a synchronised planet can be rewritten as
 \ba
 \left(\frac{de}{dt}\right)^{\rm (planet)}_{\rm (synchr)}=\,\frac{7\;}{57\;e}\, \frac{{\dot{P}\,}_{\rm orb}^{\rm (planet)}}{P_{\rm orb}}\,
 \simeq\, \frac{0.123}{e}\, \frac{{\dot{P}\,}_{\rm orb}^{\rm (planet)}}{P_{\rm orb}}\,\;,
 \label{above}
 \ea
 which includes only the observed parameters: the eccentricity, the orbital period, and its time derivative ${\dot{P}\,}_{\rm orb}=\,{\dot{P}\,}_{\rm orb}^{\rm (planet)}$, where the superscript $\rm (planet)$ serves to remind us that equation (\ref{above}) was derived under the assumption that tidal evolution is dominated by a synchronised planet. This equation is then valid for any exoplanet system for sufficiently small eccentricity and obliquity irrespective of the planet's rheology, mass, or orbital separation from the star, insofar as the planet is synchronised and the observed rate of orbital period evolution is caused by tidal dissipation within it. This expression renders a slower eccentricity decay rate for a greater current value of eccentricity and a fixed rate of period decay. The corresponding instantaneous relative rates of decay then have a simple relation:
 \ba
 \frac{\dot P_\mathrm{orb}}{P_\mathrm{orb}}\, = \,\frac{57}{7}\,e^2\,\frac{\dot e}{e}\;\,,
 \ea
 where we omitted the object-specific index, because it is also valid for the star, if it is synchronised
     and if the tides in it are dominating the orbital evolution.
 The critical value of eccentricity for which the relative rates of orbital period and eccentricity decay become equal is $0.35$. The vast majority of detected exoplanets in close orbits are believed to have much smaller values of eccentricity though. Therefore, most of the known exoplanets should evolve much faster in eccentricity than in orbital period, and this is in tension with the observed rates of orbital period decay.

 \section{Estimate of the effect of apsidal precession}

 Tidal deformations of close synchronised planets gives rise to a fast precession of apsides \citep[e.g.,][]{2009ApJ...698.1778R}. A finite orbital eccentricity makes this precession observable as a periodic variation of transit times. An alternative explanation for the observed rate of orbital period evolution may then be a fast precession of periastron coupled with a finite orbital eccentricity. Ignoring the small difference between the sidereal and anomalistic orbital periods and assuming that the orbital inclination to the line of sight equals $\pi/2$, 
 the fitting model for the transit time in the presence of periastron precession is \citep{1995Ap&SS.226...99G}
 \ba
 t_k=t_0+P_{\mathrm{orb}}\,k-\frac{e\,P_{\mathrm{orb}}}{\pi}\cos\left(\omega_0+\frac{d\omega}{dk} k\right)+O(e^2),
 \ea
 where $\omega_0$ is the argument of periastron at time $t_0$, and $k$ is the scaled time (orbit counter) equal to $t/P_\mathrm{orb}$. The cosine term is responsible for periodic variations of $\dot P_\mathrm{orb}$, so that the observed interval between consecutive transits may increase or decrease depending on the argument of periastron $\omega$ at the time of measurement. The fastest rate of period decline is achieved at $\omega_0=3\pi/2$ where the observed rate is
 \ba
 \dot P_\mathrm{orb}^{\rm \,(max)}\simeq -\;\frac{e\,P_\mathrm{orb}}{\pi}\frac{d\omega}{dt}.
 \ea
 Substituting the well-known formula for the rate of apsidal precession  \citep[e.g.,][]{2009ApJ...698.1778R}, one obtains 
 \ba
 \stackrel{\bf\centerdot}{P\,}_{\rm orb,\,prec}^{\rm (max)} \;= \qquad\qquad\qquad\qquad\qquad\qquad\qquad\qquad\qquad\quad
 \label{pmax.eq}\\
 \nonumber
 -\;15\s k_{2,p}\,e\,\left(\frac{R_{\rm p}}{a}\right)^{\textstyle{^{5}}}\frac{\;M\;}{M_{\rm p}}\;\frac{1}{(1-e^2)^5}\left(1+\frac{3}{2} e^2+\frac{1}{8} e^4\right) \,\;,
 \ea
 with $k_{2,p}$ being the planet's Love number. With the estimated values of input parameters quoted above, the measured rate of period decay can be achieved if the Love number is equal to or greater than 0.26, consistent with values inferred for Jupiter and Saturn. The value of eccentricity used in this calculation ($e=0.0628$) is taken from \citep{Chontos2019}, where it is given with a formal uncertainty of 31\%. At the lower bound uncertainty interval, the minimum required $k_{2,p}$ becomes 0.37. 
 This is therefore a plausible alternative explanation (to efficient stellar tidal dissipation) of the observed $\dot{P}_\mathrm{orb}$ for Kepler-1658\,b, but it requires a suitable orientation of the orbit and a sufficiently high value of the planet's Love number.

 In addition to the tidal deformation of the planet (and of the host star), a nearly constant oblateness of their figures is generated by rotation. Adopting a formula for the rate of apsidal precession from  \citep{2009ApJ...698.1778R}, we can write, for $M\gg M_p\;$:
\ba
\stackrel{\bf\centerdot}{P\,}_{\rm orb,\,rot}^{\rm (max)} \;=
 \label{prot.eq}
 -\s k_{2}\,e\,\left(\frac{R}{a}\right)^{\textstyle{^{5}}}\frac{\;{\dot\theta}^{\,2}\;}{n^2}\;(1-e^2)^{-2} \,\;,
 \ea
 where $k_{2}$ is the star's Love number, and $\dot\theta$ is angular frequency of rotation of the star.
In Section \ref{star.sec}, we inferred that the star rotates with a period of a few days and may be synchronised by the planet. Assuming that $\dot\theta=n$, the ratio of the maximum rates of orbital decay caused by the tidal deformation of the planet  and the rotational deformation of the star is
\ba
\frac{\stackrel{\bf\centerdot}{P\,}_{\rm orb,\,prec}^{\rm (max)}}{\stackrel{\bf\centerdot}{P\,}_{\rm orb,\,rot}^{\rm (max)}}\simeq
15\,\frac{k_{2,p}}{k_{2}}\left(\frac{R_{\rm p}}{R}\right)^5
\frac{\;M\;}{M_{\rm p}}\,(1-e^2)^{-3}\left(1+\frac{3}{2} e^2+\frac{1}{8} e^4\right).
\ea
Note that this ratio is independent of semi-major axis. Surprisingly, this equation implies that the rate of apsidal precession is dominated by the contribution due to the rotational deformation of the star. This contribution is roughly 3000 times larger than the previously estimated one due to the tidal deformation of the planet.
Within this model, the open issue is why the observed rate of orbital period decay is so low. There are a few possible routes to address it, including an unfavourable orientation of the orbit with respect to the line of sight, a vanishingly small value of orbital eccentricity, or a large value of the orbital obliquity on the equator of the star. Obviously, these scenarios also require some tuning of the configuration parameters.

 \section{Tidal dissipation in a non-synchronous planet}
 \label{Section_6}

 We now explore a different possibility, that of the planet rotating non-synchronously at the present day. Since tidal synchronisation would be expected to occur well within the age of the system, this scenario requires us to invoke an additional process. For an estimate, we use equation (\ref{equation}), having interchanged in it $M$ with $M_{\rm p}$, and having substituted $R$ and $K_2(\omega_{lmpq})$ with $R_{\rm p}$ and $K^{\,\prime}_2(\omega^{\,\prime}_{lmpq})\s$, correspondingly:
 \ba
 \stackrel{\bf\centerdot}{P\,}_{\rm orb}^{\rm (planet)}\;
 \approx\,-\,9\s\pi\,
 \frac{M\;}{M_{\rm p}}\,\left(\frac{R_{\rm p}}{a}\right)^{\textstyle{^5}}\,K^{\,\prime}_{2}(2n-2\dot{\theta}_{\rm p})
 \;\;,
 \label{}
 \ea
 $\dot{\theta}_{\rm p}$ being the planet's rotation rate. The ensuing dissipative properties of the non-synchronised planet are given by
 \ba
 \nonumber
 K_2^{\,\prime}(2n-2\dot{\theta}_{\rm p})\s
    &\approx&
 \;-\;3.54\times 10^{-2}\,\frac{M_{\rm p}}{\;M\;}\;\left(\frac{a}{R_{\rm p}}\right)^{\textstyle{^{5}}}{\stackrel{\bf\centerdot}{P\,}_{\rm orb}^{\rm (planet)}}
    ~\\   &=&
 \;8.70\times 10^{-3}\;,
 \ea
 \ba
 Q^{\,\prime}_{\rm p}=\frac{3}{2}\,\frac{1}{|K^{\,\prime}_2|}\,=\,1.72\times 10^2
 \,\;.\quad
 \label{}
 \ea
 These values predict more efficient tidal dissipation in the planet, by at least an order of magnitude, than expected for giant planets, comparing with inferences in the Solar system by \citet{2009Natur.459..957L, 2017Icar..281..286L}. These values are more appropriate for what might be anticipated of a highly dissipative hot super-Earth. It is possible that this efficient dissipation might be explained by Kepler-1685b containing a large viscoelastic core with the right properties \citep[e.g.][]{Remus2012,Storch2014}, or by it being  locked into resonance with a global mode \citep[e.g.][]{Fuller}, but both of these possibilities are highly uncertain.

 This situation is analogous to the case of the hot Jupiter WASP-12b spiralling into its host. Most models of tidal dissipation in the star WASP-12 matching observational constraints are unable to produce the measured rate of orbital decay, though this remains a possibility there, as it does for Kepler 1658, due to uncertainties in stellar models and their tidal response. The dissipation rate in a synchronously rotating WASP-12b is unable to account for this decay rate either.  At the same time, the tidal dissipation rate in a nonsynchronous WASP-12b explains the observations~---~and renders for this planet a value of $Q_p^{\,\prime}$ very close to that of our own Jupiter \citep{WASP-12b}.

 One potential explanation for the planet Kepler-1685b staying non-synchronous with respect to its orbit could be synchronisation of the planet with its own moon \citep{Pathways}. This option, however, imposes fairly tight restrictions on the properties of the putative moon. As demonstrated in Appendix \ref{Appendix_A}, for Kepler-1658\,b the required mass of the moon turns out to be prohibitively large, which makes this explanation unlikely. 
 
\section{Conclusions and further questions}

 The Kepler-1658 system currently affords us the rare opportunity to constrain the efficiency of stellar tidal dissipation by measuring tidal decay around a post-main-sequence star, albeit one which has only very recently left the main-sequence. In this investigation, we have performed a critical analysis to determine values of the modified tidal quality factor that may explain the observed orbital period decay.
 
 We employed PARSEC evolution models to obtain, from the Gaia DR3 catalog, values of the key parameters of the star Kepler-1658\,a, including constraining its age, mass $M$, radius $R$, and effective temperature $T_{\rm eff}$. Using the obtained values, we constructed interior models of this star and performed calculations to determine tidal dissipation rates predicted by several mechanisms, including both equilibrium and dynamical tides. Combining the constraints based on stellar effective temperature $T_{\rm eff}$ and radius $R$, we have concluded that the likeliest value of its modified quality factor indicated by our models is $Q^{\,\prime}\simeq 3\times 10^{9}\s$. This value is far too large to account for the measured rate of tidal decay of the planet Kepler-1658\,b. Values of $Q^{\,\prime}$ sufficient to explain observations can be obtained due to inertial waves in the convective envelope of the star based on the looser constraints on $R$ only, but not in our models that combine constraints on both $T_{\rm eff}$ and $R$. This suggests the possibility that contraction of the orbit may be due to tides in the planet instead.

 However, we have shown that tides in a synchronised Kepler-1658\,b are still insufficient to provide the observed rate of orbital decay, unless we endow the planet with an unrealistically low $Q^{\,\prime}_{\rm p}\simeq 13\s$. On the other hand, tidal dissipation in a nonsynchronous Kepler-1658\,b can potentially explain the orbital shrinking, if the planet's modified quality factor is as low as $Q^{\,\prime}_{\rm p}\simeq 170\s$. Such a value is commonly thought to be more appropriate to a highly dissipative hot super-Earth than to a gas giant planet, but it is possible this value may indicate that Kepler-1685b contains a large viscoelastic core, for example.

 As explained in Section \ref{Section_6}, a putative nonsynchronous rotation rate of the planet cannot be explained by the presence of a massive moon that might have synchronised the planet with the moon's mean motion about it. For that to happen, the moon must be abnormally massive. Still, other mechanisms preventing synchronicity are possible. For example, given the substantial planetary eccentricity, a higher spin-orbit state cannot be excluded. Another option could be thermal tides which can both push planets away from synchronism and excite their eccentricities \citep[e.g.][]{Arras2010}. Still another possibility could be differential rotation, meaning that the planet could be synchronised on average but some layers could rotate differently, thereby boosting tidal dissipation. (For example, the surface layers may be rotating differently due to the stellar heating.) While, based on our analysis, the planet (or, at least, some of its layers) should rotate nonsynchronously within these tidal decay scenarios, we acknowledge that the reason(s) for this nonsynchronism need much further study. 
 
 On the other hand, a fast apsidal precession, caused by the rotational deformation of the host star and the tidal deformation of the planet, provides a viable interpretation of the long-term curvature in the transit time variation data. This effect is proportional to the orbital eccentricity, which is not known precisely. This may cause apparent acceleration or deceleration of the observed transit frequency depending on the current orientation of the orbit with respect to the line of sight. With the nominal best-estimate parameters, we concluded that the tidal deformation of a synchronised planet may account for the estimated rate of orbital period shrinkage if its static Love number is greater than 0.26, which seems to be viable. This explanation requires the current periastron to be close to the line of sight direction. Our estimates indicate that theoretical models of periastron precession caused by the (rotational) oblateness of the star predict orders of magnitude faster rates of orbital evolution compared to the tidal deformation of the planet, unless the Love number of the star is very small. The star Kepler-1658 stands out from the population of exoplanet hosts because of its large radius and very high rate of rotation, and these properties may account for the proposed transit time variations. The puzzle to be resolved is then why is the observed rate of orbital period shrinkage so low. Possible explanations include an unfavourable alignment of the orbit and a very small orbital eccentricity.

\section*{Acknowledgments}
AJB was funded by STFC grants ST/S000275/1 and ST/W000873/1. We would like to thank the referee for their constructive and prompt report.

\section*{Data Availability}
The data underlying this article will be shared on reasonable request to the corresponding author.

\bibliographystyle{mnras}
\bibliography{references}

\appendix

\section{Kepler-1658 b cannot be synchronised by a moon}
\label{Appendix_A}

 The putative moon should remain within the niche sandwiched between the Roche radius and the reduced Hill radius. The formula for the Roche radius,
 \bs
 \ba
 \label{}
 r_{\textstyle{_{\rm Roche}}}=2.20\, R_{\textstyle{_{\rm m}}}\,\left(\frac{M_p}{M_{\textstyle{_{\rm m}}}}\right)^{{1}/{3}}\;\,,
 \ea
 can be conveniently written as
 \ba
 \label{}
 r_{\textstyle{_{\rm Roche}}}=
  5.47\times 10^{-2}
 \, R_{\textstyle{_{\rm Jup}}}\,\left(\frac{M_p}{M_{\textstyle{_{\rm Moon}}}}
 \right)^{{1}/{3}}\,\left(\frac{\rho
 _{\textstyle{_{\rm Moon}}}
 }{\rho_m}\right)^{{1}/{3}}\;\,,
 \ea
 \es
 where $\rho$ is the average density, subscript ``Moon'' refers to our Moon, and subscript ``$m$'' refers to the exomoon. For $M_p=5.88\,M_{\rm Jup}\,$, one obtains $r_{\textstyle{_{\rm Roche}}}=2.9\,R_{\textstyle{_{\rm Jup}}}$.

 Owing to equation (2) from \cite{Pathways}, the reduced Hill radius $r_H^{\,\prime}$ for
   a prograde exomoon is $6.22\s R_{\rm Jup}\s$, \,while for a retrograde moon it is $11.8\s R_{\rm J}$.
  According to formula (51) from \cite{Pathways}, the mass required for an exomoon to be, in principle, capable of synchronising
  its planet satisfies the inequality
  \ba
  M_m\s>\s M_p\s \left(\frac{R_p}{r^{\,\prime}_H}\right)^{2}\,\;.
  \label{}
  \ea
  For the Kepler-1658 b planet, this implies that the mass of a prograde moon should be at least 3\% of the planet's mass $M_p$. With the proposed value of $M_p$, the required mass of the exomoon must exceed 55 Earth masses, which is unrealistic. The mass of a retrograde moon must be at least 0.8\% $M_p$, i.e., about 14.6 Earth masses, which is still an unlikely option.

\bsp
\label{lastpage}

\end{document}